# Neutrino-bound di-neutrons as an exotic metastable atom

Yu.L. Ratis

Institute of power engineering for the special application

443071, Volgsky pr.33-87, Samara, Russia, ratis@samtel.ru

Image processing systems institute RAS

443001, Molodogvardeiskaya, 151, Samara, Russia, ipsi@smr.ru

**Abstract**

It is shown the possibility of existence of metastable atoms of dineutroneum as a bound state of two neutrons and one neutrino. Such atoms can appear in a reaction of deuterons with free or quasi-free electrons. The estimation of mass, size and lifetime of dineutroneum atom is fulfilled.

**1. Introduction**

Laws of physics do not impose basic theoretical bans on the existence of the metastable bound state of the two neutrons and neutrino [1], because a neutrino is a massive particle [2].

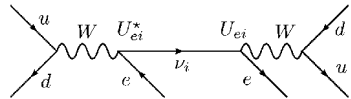

Fig. 1. The typical diagram of the electroweak process [3,4].

Due to interaction with quarks in a nucleon, a neutrino can "linger" inside it. This delay is caused, because the effective $N\nu$ - potential corresponding to $W$ - boson exchange (Fig. 1), is a short-range and very deep one. Its depth is still rather small to keep antineutrino, proton and electron in the bound state (i.e. like a neutron) for a long time, but just enough to consider a proton like the stable bound state of three particles, positron, neutron and neutrino. It is well known, that three-body effects allow an existence of 3 particles' bound states, which pair potentials are insufficiently deep to form 2 particles' bound states.

A long lifetime of the neutrino inside a nucleus can be treated on the basis of exotic Miheev – Smirnov - Volfenstein effect at low energies [5]. Let us explain this in more detail. If the energy of incoming electron is resonant (i.e. renormalized masses of all three types of neutrinos ($\nu_e, \nu_\mu, \nu_\tau$) inside a nucleon are approximately equal after the electron capture), the exotic nucleus is generated at the first stage of electroweak process (two left vertexes in the diagram 1), which cannot decay until an oscillation have been finished. The exotic nucleus $D_\nu$ is metastable, because the energy conservation law forbids its decay with $\mu$- or $\tau$- lepton emission. The channel $D_\nu \to 2n + \nu_e$ is also closed. Thus, theoretical consideration of the bound state of the neutrino inside a nucleus in the framework of any potential model gives us only phenomenological description of the observable effect.

From this standpoint, we shall consider hypothetical metastable exotic atom (exotic nucleus) dineutroneum, which is the bound state of two neutrons and one neutrino, as was mentioned above. The aim of this work is to estimate the mass, size and lifetime of the dineutroneum atom which is formed due to the interaction of deuterons with free, or quasi-free electrons.

**2. Main formalism**

The known Hamiltonian of weak interaction is

$$H' = \frac{G}{\sqrt{2}} \int J^{\lambda+}(\vec{r}) \cdot \hat{G}(\vec{r},\vec{r}') \cdot J_\lambda(\vec{r}') d\vec{r} d\vec{r}' , \qquad (1)$$

with $G$ constant of universal weak interaction Fermi, $J_\lambda(\vec{r})$ weak current, and $\hat{G}(\vec{r},\vec{r}')$ propagator. Let us introduce definition in accord to [6]

$$\begin{cases} J^{\lambda+} = (J_\lambda)^+, & \lambda = 1,2,3 \\ J^{4+} = -(J_4)^+ \end{cases}, \qquad (2)$$

and similarly for others 4- vector operators. In the standard model, the weak interaction is caused by exchange of the $W$- boson with mass $\sim 90\,GeV$. Therefore, if we consider the low energy weak processes, an approximation $m_W \to \infty$ can be used. Accordingly, the interaction is quite local, and components of the weak current in Hamiltonian (1) should be taken at the same point of space ($\hat{G}(\vec{r},\vec{r}') = \delta(\vec{r} - \vec{r}')$). Hence

$$H' = \frac{G}{\sqrt{2}} \int J^{\lambda+}(\vec{r}) \cdot J_\lambda(\vec{r}) d\vec{r} . \tag{3}$$

The Lorenz invariant weak current is well known. For example, $\beta$ - decay of a neutron is described by the Hamiltonian [6]

$$H' = \frac{G}{\sqrt{2}} \int \left[\overline{\psi}_n(\vec{r})\gamma^\lambda(1+\gamma_5)\psi_p(\vec{r})\right]^+ \cdot \left[\overline{\psi}_e(\vec{r})\gamma_\lambda(1+\gamma_5)\psi_{\nu_e}(\vec{r})\right] d\vec{r} . \tag{4}$$

To describe the weak processes in nuclear physics, one needs in a non-relativistic Hamiltonian $h'(\vec{r})$. The model of the Hamiltonian was derived in the early papers of Fermi, Gamov and Teller, and looks like [6]

$$h'(\vec{r},t) = \frac{G}{\sqrt{2}} \left\{ i\beta [f_1\gamma_\lambda + f_2\sigma_{\lambda\rho}k_\rho + (g_1\gamma_\lambda + ig_2 k_\lambda)\gamma_5] \right\}^+ j^\lambda(\vec{r},t) + h.c. \tag{5}$$

In (5)

$$j^\lambda(\vec{r},t) = \left[ i\overline{\psi}_l(\vec{r})\gamma^\lambda (1+\gamma_5)\psi_{\nu_l}(\vec{r}) \right] \cdot \exp\left(-\frac{i}{\hbar}(E_{\nu_l} - E_l)t\right) \tag{6}$$

the lepton current, $E$ the energy positive for particles and negative for antiparticles, $f_1$, $f_2$, $g_1$, $g_2$ the formfactors, $\psi(\vec{r})$ lepton wave function (WF).

In the works devoted to the nuclear $\beta$ - processes, the WFs of free leptons in (6) are usually chosen as plane waves with the momentum $\vec{p}_l$[1]. Thus, the lepton's current (6) looks like:

$$j_\lambda(\vec{r},t) = L^{-3} b_\lambda \exp(i\vec{k}\cdot\vec{r}) \cdot \exp\left(-\frac{i}{\hbar}(E_\nu - E_e)t\right) \tag{7}$$

where $\vec{k} = \vec{\nu} - \vec{e}$ the lepton transferred momentum, $\vec{\nu}$ the wave vector of neutrino, $\vec{e}$ the wave vector of electron, $L^3$ the normalization volume,

$$b_\lambda(\underline{m}_e, \underline{m}_\nu) = i\left(\overline{u}_e(\underline{m}_e)\gamma_\lambda w_\nu(\underline{m}_\nu)\right), \tag{8}$$

and

$$w_\nu(\underline{m}_\nu) = (1+\gamma_5)u_\nu(\underline{m}_\nu) . \tag{9}$$

The spinor

$$w_\nu(\underline{m}_\nu) = \frac{1}{\sqrt{2}}\begin{pmatrix}1\\-1\end{pmatrix} \cdot \left(1 - (\vec{\sigma}\cdot\hat{\vec{\nu}})\right) \cdot \chi_{1/2}(\underline{m}_\nu) \quad , \tag{10}$$

---

[1] In reactions of electron capture, $\beta$ - decay into a bound state and in mesoatoms the charged lepton occupies the bound state and its WF belongs to the discrete spectrum.

$m_\nu = \pm 1/2$ the spin projection of neutrino (+ corresponds to spin "up" and − spin "down").

The lifetime of dineutroneum can be estimated within the approximation of allowed transitions. Therefore, we shall neglect the small contribution of the terms $\hbar k/(Mc)$, $p/Mc$ и $kR$ due to the forbidden transitions, and obtain the non-relativistic limit of the Hamiltonian (5) in the plane wave approximation:

$$h'(\vec{r}) = \frac{G}{\sqrt{2}\cdot L^3} e^{i\vec{k}\cdot\vec{r}} \cdot \sum_{j=1}^{A}\left[if_1\cdot b_4 - g_1\left(\vec{b}\cdot\vec{\sigma}\right)\right]_j \cdot (\tau_+)_j \cdot \delta(\vec{r}-\vec{r}_j) + ... \qquad (11)$$

The Pauli matrixes $\tau_1$ and $\tau_2$ ($\tau_{+1}$, $\tau_{-1}$) are well known:

$$\begin{cases} \tau_+ = (\tau_1 + i\tau_2)/2 = -\tau_{+1}/\sqrt{2} \\ \tau_- = (\tau_1 - i\tau_2)/2 = \tau_{-1}/\sqrt{2} \end{cases} \Rightarrow \begin{cases} \tau_+|p\rangle = 0;\ \tau_+|n\rangle = |p\rangle \\ \tau_-|n\rangle = 0;\ \tau_-|p\rangle = |n\rangle \end{cases}. \qquad (12)$$

The approximated Hamiltonian (11) is used to describe the nuclear processes with the dineutroneum.

First, we take into account, that mass of dineutroneum is less, than double mass of the neutron. Therefore, neutrino in the atom of dineutroneum is in the bound state, and the Hamiltonian looks like

$$h'(\vec{r}) = \frac{G_\beta}{\sqrt{2}\cdot L^{3/2}} \psi_\nu(\vec{r}_c)\cdot e^{-i\vec{e}\cdot\vec{r}} \cdot \left\{\sum_{i=1}^{2}\delta(\vec{r}-\vec{r}_i)\left[ib_4 - \lambda\cdot(\vec{b}\cdot\vec{\sigma}^{(i)})\right]\tau_+^{(i)}\right\} + h.c., \qquad (13)$$

$\psi_\nu(\vec{r}_c)$ the spatial part of the neutrino's WF, $G_\beta = f_1 \cdot G$, index $c$ indicates the radius-vector of the neutrino which origin is in the centre-mass of the dineutroneum because of translation-invariance of the Hamiltonian $h'(\vec{r})$.

According to a "golden Fermi's rule", the probability of the transition to the continuum states per unit of time is equal:

$$dw_{fi} = \frac{2\pi}{\hbar}\cdot\delta\left(E_f - E_i\right)\cdot\left|\langle f|V|i\rangle\right|^2 dn_f. \qquad (14)$$

Hence, the decay probability of the bound state of two neutrons and one neutrino within the channel $D_\nu \to d + e^-$ per the time unit is equal:

$$w_{D_\nu \to d+e^-} = \frac{2\pi}{\hbar}\int \frac{L^3 d\vec{p}_e}{(2\pi\hbar)^3}\cdot\frac{L^3 d\vec{p}_d}{(2\pi\hbar)^3}\cdot\delta(E_i - E_f)\cdot\left\langle\left|\int\left\langle d\left|h'(\vec{r}')\right|D_\nu^{(N)}\right\rangle d\vec{r}'\right|^2\right\rangle. \qquad (15)$$

The WFs $\left|D_\nu^{(N)}\right\rangle$ and $\langle d|$ depend on the coordinates, spins and isospins of nucleons, and matrix elements of the transition $D_\nu \to d + e^-$ in the space of leptons are already included into the Hamiltonian $h'(\vec{r})$ by definition. The external triangular brackets in (15) mean the averaging by projections of spins of all initial particles, and analogous summation in the final state.

Let us now consider the $\beta$- decay of the dineutroneum. The initial and final states in this case are[2]:

$$\begin{cases} \left|D_\nu^{(N)}\right\rangle = \frac{1}{\sqrt{L^3}} e^{i\vec{k}_{D_\nu}\vec{R}_{D_\nu}} \psi_{2n}(\vec{r}_2 - \vec{r}_1)\chi_{00}(\vec{S})\chi_{1-1}(\vec{T}) \\ |d\rangle = \frac{1}{\sqrt{L^3}} e^{i\vec{k}_d\vec{R}_d} \psi_d(\vec{r}_2 - \vec{r}_1)\chi_{1m_d}(\vec{S})\chi_{00}(\vec{T}) \end{cases} \quad (16)$$

Consequently, the matrix element in (15) looks like

$$\int \left\langle d \left| h'(\vec{r}') \right| D_\nu^{(N)} \right\rangle d\vec{r}' = \\ = \frac{1}{L^3} \int d\vec{r}' d\vec{r}_1 d\vec{r}_2 e^{i\left(\vec{k}_{D_\nu}\vec{R}_{D_\nu} - \vec{k}_d\vec{R}_d\right)} \psi_d^*(\vec{r}'')\psi_{2n}(\vec{r}'') \left\langle \chi_{1m_d}^+(\vec{S})\chi_{00}^+(\vec{T}) \left| h'(\vec{r}') \right| \chi_{00}(\vec{S})\chi_{1-1}(\vec{T}) \right\rangle \quad (17)$$

where $\vec{r}'' = \vec{r}_2 - \vec{r}_1$.

The "nuclear" spin of the dineutroneum $J_i = 0$ and the deuteron's spin $J_f = 1$. Thus, we deal with the Gamov - Teller transition. According to it

$$h'_{GT}(\vec{r}) = \frac{-\lambda \cdot G_\beta}{\sqrt{2} \cdot L^{3/2}} \psi_\nu(\vec{r}_c) \cdot e^{-i\vec{e}\cdot\vec{r}} \cdot \left\{ \sum_{i=1}^{2} \delta(\vec{r} - \vec{r}_i) \cdot (\vec{b}\cdot\vec{\sigma}^{(i)}) \cdot \tau_+^{(i)} \right\} + h.c. \quad (18)$$

We consider the dineutroneum $\beta$- decay in its rest system. In this case $k_{D_\nu} = 0$, and (18) is simplified (details see in the Appendix):

$$\int \left\langle d \left| h'(\vec{r}') \right| D_\nu^{(N)} \right\rangle d\vec{r}' = \frac{\lambda \cdot G_\beta \sqrt{3}}{2L^{9/2}} C_{1-m_d\ 1/2\ m_\nu}^{1/2\ m_e} \int d\vec{R} d\vec{r} e^{-i\vec{k}_d\vec{R}} \psi_d^*(\vec{r})\psi_{2n}(\vec{r}) \sum_{i=1}^{2} \psi_\nu(\vec{r}_i - \vec{R}) \cdot e^{-i\vec{e}\cdot\vec{r}_i} \quad (19)$$

We determine the formfactor

$$f_{overlap}^{d \Leftrightarrow D_\nu^{(N)}}(|\vec{e}|) = \int \cos(\vec{e}\cdot\vec{r}/2) \psi_d^*(\vec{r}) \psi_\nu(\vec{r}/2) \psi_{2n}(\vec{r}) d\vec{r} \equiv \left(V_{eff}^{D_\nu}\right)^{-1/2}. \quad (20)$$

The $V_{eff}^{D_\nu}$ means an effective volume of exotic atom of dineutroneum. This circumstance allows to present eq. (19) in the extremely compact form:

---

[2] See details in [17].

$$\int \langle d | h'(\vec{r}') | D_\nu^{(N)} \rangle d\vec{r}' = \frac{\lambda \cdot G_\beta \sqrt{2}}{L^{9/2}} (2\pi)^3 \delta(\vec{k}_d + \vec{e}) f_{overlap}^{d \Leftrightarrow D_\nu^{(N)}}(|\vec{e}|)(-1)^{1/2+m_\nu} C_{1/2-m_e\ 1/2 m_\nu}^{1 m_d}. \tag{21}$$

In turn, eq. (15) can be presented in the form which is suitable for numerical calculations

$$w_{D_\nu \to d+e^-} = \frac{2\pi}{\hbar} \int \frac{d\vec{p}_e}{(2\pi\hbar)^3} \cdot \delta(E_i - E_f) \cdot 3 \cdot \left| \lambda G_\beta f_{overlap}^{d \Leftrightarrow D_\nu^{(N)}}(|\vec{e}|) \right|^2, \tag{22}$$

and evaluate the integral

$$I_{D_\nu \to d+e^-}^{ph}(p_e) = \int d\vec{p}_e \cdot \delta(E_i - E_f) = 4\pi \int dp_e p_e^2 \delta(E_{D_\nu} - E_d - E_e) \tag{23}$$

All the particles in our case are non-relativistic. Consequently,

$$\begin{cases} E_{D_\nu} = m_{D_\nu} c^2 + \dfrac{p_{D_\nu}^2}{2m_{D_\nu}} \\ E_d = m_d c^2 + \dfrac{p_e^2}{2m_d} \\ E_e = m_e c^2 + \dfrac{p_e^2}{2m_e} \end{cases}. \tag{24}$$

As a result,

$$I_{D_\nu \to d+e^-}^{ph}(p_e) \approx 4\pi p_e m_e \tag{25}$$

where the momentum

$$p_e = \sqrt{2m_e \left(m_{D_\nu} c^2 - m_d c^2 - m_e c^2\right)}, \tag{26}$$

corresponds to $\vec{p}_{D_\nu} = 0$ in the rest system of dineutroneum.

The <u>internal</u> energy of the dineutroneum $U_{D_\nu}$ is equal

$$U_{D_\nu} = m_{D_\nu} c^2 - m_d c^2 - m_e c^2 > 0 \tag{27}$$

Thus, eq. (26) can be presented in rather compact form

$$p_e = \sqrt{2m_e U_{D_\nu}}, \tag{28}$$

and we get the following expression:

$$w_{D_\nu \to d+e^-} = \frac{3}{\pi\hbar^4} \cdot m_e \cdot \sqrt{2m_e U_{D_\nu}} \cdot \left| \lambda G_\beta f_{overlap}^{d \Leftrightarrow D_\nu^{(N)}}(|\vec{e}|) \right|^2. \tag{29}$$

The momentum dependence of the formfactor (20) at the low energies can be neglected

$$f_{overlap}^{d \Leftrightarrow D_\nu^{(N)}} = \int \psi_d^*(\vec{r}) \psi_\nu(\vec{r}/2) \psi_{2n}(\vec{r}) d\vec{r} \equiv \left(V_{eff}^{D_\nu}\right)^{-1/2}, \tag{30}$$

and

$$w_{D_\nu \to d+e^-} = \frac{3|\lambda|^2 \cdot |G_\beta|^2}{\pi \hbar^4 V_{eff}^{D_\nu}} \cdot m_e \cdot \sqrt{2m_e U_{D_\nu}}.  \tag{31}$$

Formula (30) determines the overlap integral $f_{overlap}^{d \Leftrightarrow D_\nu^{(N)}}$. For estimations we accept, that the bound particles participating in the reaction $D_\nu \to d + e^-$ have the orbital momentum equal to zero, and their wave functions look like

$$\psi_d(\vec{r}) = \frac{1}{\sqrt{4\pi}} \frac{\chi_d(r)}{r}; \quad \psi_{2n}(\vec{r}) = \frac{1}{\sqrt{4\pi}} \frac{\chi_{2n}(r)}{r}; \quad \psi_\nu(\vec{r}) = \frac{1}{\sqrt{4\pi}} \frac{\chi_\nu(r)}{r}. \tag{32}$$

Only Hulten's WF $\psi_d(\vec{r})$ in (32) is known

$$\chi_d(r) = A_d \exp(-\alpha_d r)[1 - \exp(-\mu r)] \tag{33}$$

with the normalization constant

$$A_d = [2\alpha_d(\alpha_d + \mu)(2\alpha_d + \mu)]^{1/2} \mu^{-1}. \tag{34}$$

Here $\alpha_d = \sqrt{m_N |E|}/\hbar \approx 0.232\, fm^{-1}$, $\mu \approx 1.1\, fm^{-1}$ [7].

We assume that

$$\chi_{2n}(r) = A_{2n} \exp(-\alpha_{2n} r)[1 - \exp(-\mu r)], \tag{35}$$

with

$$A_{2n} = [2\alpha_{2n}(\alpha_{2n} + \mu)(2\alpha_{2n} + \mu)]^{1/2} \mu^{-1}, \tag{36}$$

and equal parameters $\mu$ for deuteron and dineutroneum.

For the sake of simplicity we suppose

$$\chi_\nu(r) = A_\nu \exp(-2\kappa r), \tag{37}$$

where $A_\nu = [4\kappa]^{1/2}$.

According to (30)

$$f_{overlap}^{d \Leftrightarrow D_\nu^{(N)}} = \frac{2}{\sqrt{4\pi}} \int_0^\infty \frac{\chi_d(r)\chi_\nu(r/2)\chi_{2n}(r)}{r} dr. \tag{38}$$

This integral in a view of (33), (35) and (37) can be calculated analytically

$$f_{overlap}^{d \Leftrightarrow D_\nu^{(N)}} = \frac{A_{2n} A_d A_\nu}{\sqrt{\pi}} \ln\left(\frac{(\alpha_{2n}^{(\nu)})^2}{(\alpha_{2n}^{(\nu)})^2 - \mu^2}\right) \tag{39}$$

where $\alpha_{2n}^{(\nu)} = \kappa + \alpha_d + \alpha_{2n} + \mu$. In this work, we suppose $\chi_d(r) \approx \chi_{2n}(r)$ (i.e. $\alpha_{2n} \sim \alpha_d$).

Let us estimate $V_{eff}^{D_\nu}$ in the rough approximation $\alpha_{2n} = \alpha_d$. The decaying dineutroneum is created in the reaction of electron capture by deuteron. Thus, we suppose neutrino to be "smeared" in a deuteron. This assumption implies an estimation $\kappa = \alpha_{2n} = 0.232\ fm^{-1}$. Consequently, we estimate $V_{eff}^{D_\nu} \approx 20\ fm^3$.

The standard Coulomb corrections also can be considered

$$w_{D_\nu \to d + e^-} = \frac{3|\lambda|^2 \cdot |G_\beta|^2}{\pi \hbar^4 V_{eff}^{D_\nu}} \cdot m_e \cdot p_e \cdot F(\eta). \qquad (40)$$

The Fermi function $F(\eta)$ in the "point-like deuteron" approximation is equal [8]

$$F(\eta) \approx \pi\eta \cdot \exp(\pi\eta) \operatorname{sh}^{-1}(\pi\eta). \qquad (41)$$

All previous calculations were carried out under the assumption, that neutrino inside the dineutroneum is the electron's neutrino $|\nu_e\rangle$. Taking account the MSV- effect, we insert the electron's neutrino weight $\langle |\langle \nu | \nu_e \rangle|^2 \rangle \sim \frac{1}{2} \div \frac{1}{3}$ into (40) [9]:

$$w_{D_\nu \to d + e^-} = \langle |\langle \nu | \nu_e \rangle|^2 \rangle \cdot \frac{3|\lambda|^2 \cdot |G_\beta|^2}{\pi \hbar^4 V_{eff}^{D_\nu}} \cdot m_e \cdot p_e \cdot F(\eta), \qquad (42)$$

where $\langle |\langle \nu | \nu_e \rangle|^2 \rangle$ is the probability for the neutrino to be in the state $|\nu_e\rangle$ in the dineutroneum.

In the table 1 the values of $w^0$, $w^c$ and a lifetime $\tau_{D_\nu}^c = 1/w_{D_\nu \to d + e^-}^c$ are displayed. An approximation $V_{eff}^{D_\nu} = 20\ fm^3$, $\langle \nu | \nu_e \rangle = 1$ is used.

| $T_e\ [eV]$ | $w^0_{D_\nu \to d + e^-}$ | $w^c_{D_\nu \to d + e^-}$ | $\tau^c_{D_\nu}$ |
|---|---|---|---|
| 0.1 | 16.5 | $1.1 \cdot 10^3$ | $9.3 \cdot 10^{-4}$ |
| 1.0 | $4.8 \cdot 10^1$ | $1.1 \cdot 10^3$ | $9.3 \cdot 10^{-4}$ |
| 10 | $1.5 \cdot 10^2$ | $1.1 \cdot 10^3$ | $9.3 \cdot 10^{-4}$ |
| $10^2$ | $4.8 \cdot 10^2$ | $1.2 \cdot 10^3$ | $8.3 \cdot 10^{-4}$ |
| $10^3$ | $1.5 \cdot 10^3$ | $2.1 \cdot 10^3$ | $4.7 \cdot 10^{-4}$ |

Table 1

The energy dependence of $w^0$, $w^c$ and the lifetime $\tau^c_{D_\nu}$

We can see from the table 1, that at the low energies the rate of the $\beta$- decay of the dineutroneum can increase almost by two orders of magnitude owing to the Coulomb interaction. At $T_e > 1\ KeV$ this effect becomes insignificant.

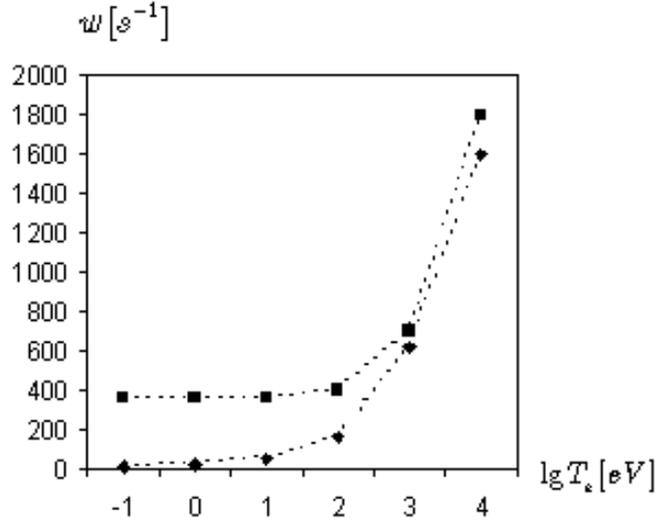

Fig. 2. The energy dependence of $w^0$, $w^c$. $V_{eff}^{D_\nu} = 20\ fm^3$. $\left\langle \left|\langle \nu | \nu_e \rangle\right|^2 \right\rangle = 1/3$.
♦ without, and ■ with the Fermi factor $F(\eta)$.

Fig. 2. demonstrates the influence of the Coulomb attraction of $\beta$- electron and deuteron-target which results in almost complete energy independence of $w = w(T_e)$ at $T_e \leq 1 KeV$. Therefore, if the dineutroneum atom is created, it lives long enough. The threshold of its creation is estimated at the level $10 - 15\,eV$, what is much lower than that for thermonuclear reactions $T_{tresh} \ll T_{tn} \sim 10\ KeV$.

Let us consider the dependence of the dineutroneum lifetime on its size. This dependence should be taking into account, since the triplet length of the neutron-neutron scattering much exceeds the deuteron's effective radius $r_d$. Table 2 demonstrates the results of theoretical calculations of the $\beta$- decay rate $w^c_{D_\nu \to d+e^-}$ and lifetime $\tau^c_{D_\nu}$ as a function of the parameter $\alpha_d / \alpha_{2n}$ at $T_e = 10\ eV$ ($\kappa = \alpha_{2n}$).

| $\alpha_d / \alpha_{2n}$ | $w^c_{D_\nu \to d+e^-}$ | $\tau^c_{D_\nu}$ |
|---|---|---|
| 1 | $1.1 \cdot 10^3$ | $9.3 \cdot 10^{-4}$ |
| 10 | $3.1 \cdot 10^3$ | $3.2 \cdot 10^{-2}$ |
| $10^2$ | $3.6 \cdot 10^{-1}$ | 2.7 |

Table 2

The dependence of rate of the $\beta$- decay of the dineutroneum on the ratio $\alpha_d / \alpha_{2n}$.

It follows from Table 2, that if the size of dineutroneum alike the size of deuterium mesoatom, its lifetime would be almost 3 seconds. Consequently, one can conclude that the exotic dineutroneum atom is metastable and its lifetime $\tau_{D_\nu} \sim 10^{-3} \, s$, i.e. three orders more than lifetime of the muon [2] $\tau_\mu = (2.197019 \pm 0.000021) \cdot 10^{-6} \, s$.

Our preliminary analysis shows, that the properties of dineutroneum: metastability, electrical neutrality and the small sizes, allow nuclear reactions of dineutroneum with nuclei in condensed matter. If we take into account large cross section of $e$ - capture ($\sigma \sim 10 \, mbarn$ for the $e^- + D \to D_\nu + X$ reaction [1]), it is possible easily explain a numerous experimental data on cold fusion in the condensed matter (see [1]). For example, there are observed [9,10] such reactions as

$$D_\nu + {}^A_{46}Pd \to {}^{A+1}_{46}Pd + p + e^-, \qquad (43)$$

$$D_\nu + p \to \begin{cases} t + \nu_e + 5.45 \; MeV \\ {}^3_2He + e^- + 5.47 \; MeV \end{cases} \qquad (44)$$

and

$$D_\nu + d \to {}^4_2He + e^- + 23.85 \; MeV. \qquad (45)$$

Thus, there is no neutrons' emission, but a creation of helium-3, helium-4 and tritium, and changed natural abundance ratio of palladium isotopes, and a plenty of excess energy. These effects were observed in the experiments [9,10] (see Fig. 3).

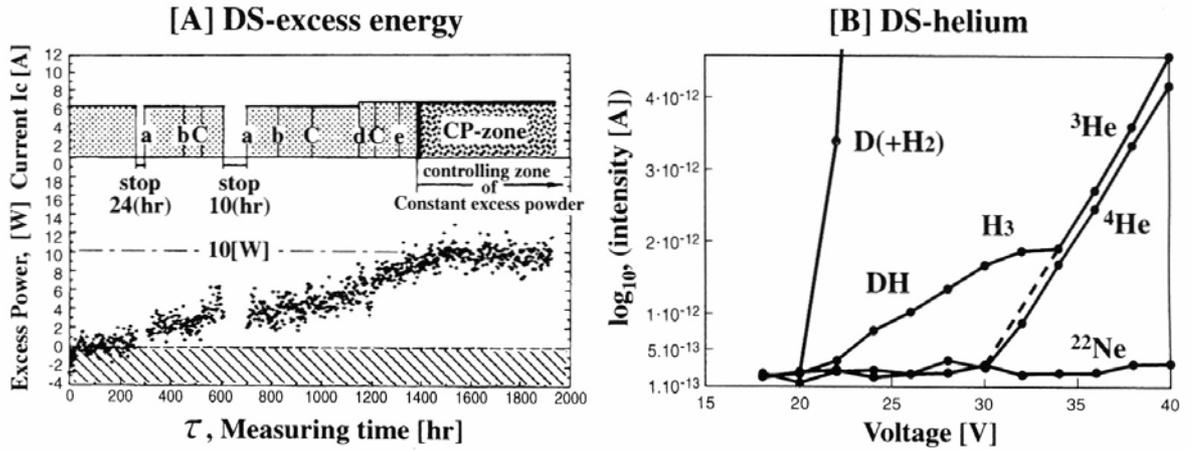

Figure 3. Excess energy and helium (3He, 4He) created inside "open type DS-cathode" which can measure continuously the change of inner pressure inside DS-cathode, and called shortly "DS-excess energy" and "DS-helium", respectively. [A] DS-excess energy. [B] DS-helium using "Vi-effect". Left panel, (a) current stop; (b) inner gas test; (c) inner gas test; (d) added 20 cc; (e) added 10 cc, renewed 20 cc; (CP) controlling zone to get constant excess power [9] (see details in [9]).

**Conclusions:**

1. The atom of dineutroneum is metastable ($\tau_{D_\nu} \sim 10^{-3}\, s$).

2. The size of dineutroneum are commensurable with the size of deuteron.

3. The mass of dineutroneum $M_{D_\nu} = 2.014102236\, e = 1876.0979650\, MeV$.

4. Metastability, electrical neutrality and small size allow nuclear reactions of the dineutroneum exotic atoms with nuclei both in gases, and in a condensed matter (for example: $D_\nu + p \rightarrow t + \nu_e$, $D_\nu + p \rightarrow {}^{3}_{2}He + e^-$, $D_\nu + d \rightarrow {}^{4}_{2}He + e^-$). This presents the clear explanation of many experiments [9-13].

5. All these results are in a full agreement with previous estimations [14-16].

**Acknowledgments**

An author would like to thank Dr. Yu.V. Popov and Prof. F.A. Gareev for the fruitful discussions.

## Appendix.  Spin and isospins matrix elements

The isospins matrix element is equal

$$\left\langle \chi_{00}(\vec{T}) \left| \tau_-^{(i)} \right| \chi_{1-1}(\vec{T}) \right\rangle = \frac{1}{\sqrt{2}} \left\langle [p(1)n(2) - p(2)n(1)] \left| \tau_+^{(i)} \right| n(1)n(2) \right\rangle = \frac{(-1)^{i-1}}{\sqrt{2}} \qquad (1)$$

The spin matrix element is more complicated

$$S_{(i)} \equiv \left\langle \chi_{1m_d}(\vec{S}) \left| \sigma_k^{(i)} \right| \chi_{00}(\vec{S}) \right\rangle =$$
$$= \sum_{m_1,m_2} C_{1/2\,m_1\,1/2\,m_2}^{1\,m_d} \sum_{m_3,m_4} C_{1/2\,m_3\,1/2\,m_4}^{0\,0} \left\langle \chi_{1/2\,m_1}^{(1)} \chi_{1/2\,m_2}^{(2)} \left| \sigma_k^{(i)} \right| \chi_{1/2\,m_3}^{(1)} \chi_{1/2\,m_4}^{(2)} \right\rangle \qquad (2)$$

According to the Clebsh - Gordan coefficients' properties

$$\left\langle \chi_{1m_d}(\vec{S}) \left| \sigma_k^{(1)} \right| \chi_{00}(\vec{S}) \right\rangle = - \left\langle \chi_{1m_d}(\vec{S}) \left| \sigma_k^{(2)} \right| \chi_{00}(\vec{S}) \right\rangle, \qquad (3)$$

and we obtains

$$S_{(i)} = (-1)^{i-1} \sum_{m_1,m_2} C_{1/2\,m_1\,1/2\,m_2}^{1\,m_d} \sum_{m_3,m_4} C_{1/2\,m_3\,1/2\,m_4}^{0\,0} \delta_{m_2 m_4} \left\langle \chi_{1/2\,m_1}^{(1)} \left| \sigma_k^{(1)} \right| \chi_{1/2\,m_3}^{(1)} \right\rangle \qquad (4)$$

It is evident, that

$$\sigma_\mu \chi_{1/2\sigma} = -\sqrt{3} \cdot \sum_{\sigma'} C_{1\mu\,1/2\sigma}^{1/2\sigma'} \cdot \chi_{1/2\sigma'} \qquad (5)$$

Thus

$$S_{(i)} = (-1)^{i-1} \sqrt{3} \sum_{m_1,m_2,m_3} C_{1/2\,m_2\,1/2\,m_1}^{1\,m_d} C_{1/2\,m_2\,1/2\,m_3}^{0\,0} C_{1k\,1/2\,m_3}^{k\,m_1} \qquad (6)$$

and:

$$S_{(i)} = (-1)^{i-1} \sqrt{3} \cdot \sum_{m'',\sigma,\sigma'} C_{j''m''\,1/2\sigma'}^{j'm'} C_{j''m''\,1/2\sigma}^{jm} C_{1\mu\,1/2\sigma}^{1/2\sigma'} \equiv (-1)^{i-1} \sqrt{6 \cdot \hat{j}' \cdot \hat{j}} \cdot F_{ang} =$$
$$= (-1)^{i-1} \sqrt{6 \cdot \hat{j}' \cdot \hat{j}} \cdot \sum_{m'',\sigma,\sigma'} (-1)^{l-1/2+m'+j''-1/2+m+1/2+\sigma'} \begin{pmatrix} j'' & 1/2 & j' \\ m'' & \sigma' & -m' \end{pmatrix} \begin{pmatrix} j'' & 1/2 & j \\ m'' & \sigma & -m \end{pmatrix} \begin{pmatrix} 1 & 1/2 & 1/2 \\ \mu & \sigma & -\sigma' \end{pmatrix} \qquad (7)$$

where $\hat{j} \equiv 2j+1$. The sum of three $3jm$ - Wigner symbols $F_{ang}$ is equal [17]

$$F_{ang} = (-1)^{j-1/2+l+j'+m'} \begin{pmatrix} j' & 1 & j \\ m' & -\mu & -m \end{pmatrix} \begin{Bmatrix} j' & 1 & j \\ 1/2 & 1 & 1/2 \end{Bmatrix}. \qquad (8)$$

Inserting (8) in (7), we get the value $S$:

$$S_{(i)} = (-1)^{i-1} \sqrt{6(2j+1)} \cdot (-1)^{j+1/2} \cdot C_{1\mu\,jm}^{j'm'} \cdot \begin{Bmatrix} j' & 1 & j \\ 1/2 & 1 & 1/2 \end{Bmatrix}. \qquad (9)$$

Thus, we derive the result

$$\left\langle \chi_{1m_d}(\vec{S}) \left| \sigma_k^{(i)} \right| \chi_{00}(\vec{S}) \right\rangle = (-1)^{i-1} \delta_{-k,\underline{m}_d}. \qquad (10)$$